\documentclass[a4paper]{jpconf}
\usepackage{graphicx}
\begin{document}
\title{A Monte Carlo algorithm for sampling rare events: 
application to a search for the Griffiths singularity}

\author{Koji Hukushima$^1$ and Yukito Iba$^2$}
\address{$^1$
Department of Basic Science, University of Tokyo, 
3-8-1 Komaba, Meguro-ku, Tokyo 153-8902, Japan
}
\address{$^2$
Department of Statistical Modeling, 
The Institute of Statistical Mathematics, 
4-6-7 Minami-Azabu, Minato-ku, 
Tokyo 106-8569, Japan 
}

\ead{hukusima@phys.c.u-tokyo.ac.jp}


\begin{abstract}
We develop a recently proposed importance-sampling Monte Carlo algorithm
 for sampling rare events and quenched variables in random disordered
 systems. We apply it to a two dimensional bond-diluted Ising model and
 study the 
 Griffiths singularity which is considered to be due to the existence of
 rare large clusters. 
It is found that the distribution of the inverse susceptibility has an
 exponential tail down to the origin which is considered the
 consequence of the Griffiths singularity. 
\end{abstract}

\section{Introduction}
\label{sec:Introduction}

Monte Carlo (MC) methods are general computational techniques  for
sampling from a given probability distribution and for estimating
expectation values under the distribution. They have been used in a wide
range of physics \cite{BinderLandau} and statistical science \cite{Liu}. 
Most of the MC methods are based on the Metropolis importance-sampling
strategy, in which a Markov chain is constructed so that its stable
distribution converges to a required distribution. 
Some improvements on the Metropolis MC algorithm have been made in order
to accelerate the sampling process. One of them is extended ensemble MC
methods \cite{IbaE} which includes multicanonical method \cite{Berg},
simulated tempering \cite{ST} and exchange MC method \cite{EMC}(parallel
tempering).  These methods have been considered to be quite useful for
studying complex systems such as spin
glasses \cite{BinderYoung,Mezard,Young,Kawashima03}. Among them, the
multicanonical method successfully applied to statistical-mechanical
models where rare events play an important role in the physical
nature. Such examples of rare events are a mixed phase at a first-order
phase transition temperature of the two-dimensional 10-state Potts
model \cite{Berg} and tails of an order-parameter distribution of the
two-dimensional Ising model \cite{Hilfer03}.   

Another  class of rare events, which has not extensively studied yet,
is found in quenched disordered systems 
when a distribution of a physical quantity over the quenched disorder  
is taken into account. 
A typical example is the rare-event tails of the ground-state-energy
distribution of 
quenched disordered system that has been the subject of recent
theoretical studies \cite{Bouchaud,Katzgraber2005}. It is, however,
difficult for a 
conventional simple-sampling method to evaluate the tails of distribution
precisely. 
Recently K\"{o}rner, Katzgraber and Hartmann \cite{Keorner2006} have
proposed an importance-sampling MC algorithm in order to evaluate the
tails of the 
ground-state-energy distribution with high precision. Using the
algorithm they could evaluate the tails of the distribution of a
spin-glass model up to 18 orders of magnitude.

A Griffiths 
singularity \cite{Griffiths}  of random spin models is also caused by
the rare events, namely the existence of arbitrary large clusters. 
Although a dynamical aspect of the Griffiths singularity has been
confirmed by numerical simulations, there is, to our knowledge, no
numerical evidence of the singularity in static observables and the
function form
that the Griffiths singularity takes has not been well established.  
It is the
main purpose of the present work to detect numerically the Griffiths
singularity. We develop the importance-sampling algorithm mentioned
above and apply it to a finite-temperature calculation of the
susceptibility distribution of a bond-diluted Ising model which is
expected to exhibit the Griffiths singularity.  

The outline of this paper is as follows. In section~\ref{sec:Algorithm} we
explain the Monte Carlo algorithm for efficiently sampling the quenched
variables of the disordered systems. In
section~\ref{sec:Griffiths} we briefly review the Griffiths singularity in randomly
bond-diluted Ising model. In section~\ref{sec:Application} we study the
Griffiths singularity by applying the method described in
section~\ref{sec:Algorithm}. 
In section~\ref{sec:conclusion} we finally draw our conclusions and
perspectives. 

\section{Monte Carlo algorithms for quenched variables}
\label{sec:Algorithm}

A quenched disordered system is defined by a Hamiltonian
$H(\mathbf{S}|\mathbf{J})$, where $\mathbf{S}$ denotes a configuration
of the system and $\mathbf{J}$ a set of quenched variables generated by
the probability distribution $P(\mathbf{J})$. 
In spin-glass models, for instance, $\mathbf{S}$ and $\mathbf{J}$ 
correspond to the spin variables and the interaction bonds,
respectively. 
One needs to take double averages for evaluating a physical quantity
$A(\mathbf{S},\mathbf{J})$ 
in the quenched disordered system. Firstly, a thermal average $\langle
A\rangle(\mathbf{J})$ at temperature $T=1/\beta$ is calculated by
tracing over the variable 
$\mathbf{S}$  for a fixed set of $\mathbf{J}$, which is expressed as
\begin{equation}
 \langle A\rangle(\mathbf{J}) = \frac{1}{Z(\mathbf{J})}\mathrm{Tr}_{\mathbf{S}}
  \exp\left(-\beta H(\mathbf{S}|\mathbf{J})\right), 
\end{equation}
where the normalization factor $Z(\mathbf{J})$ is the partition function. The other
average is to take an average $\langle A\rangle(\mathbf{J})$ over the
distribution $P(\mathbf{J})$, 
\begin{equation}
 [\langle A\rangle] = \int d\mathbf{J} P(\mathbf{J})\langle
  A\rangle(\mathbf{J}). 
\end{equation}
In general, the distribution of the quantity $A$ is defined by 
\begin{equation}
 P(A) = \int d\mathbf{J} P(\mathbf{J}) \delta \left(A-\langle
  A\rangle(\mathbf{J}) \right),  
\label{eqn:PofA}
\end{equation}
where $\delta$ is the Dirac delta function. The aim of this work is to
estimate accurately the distribution $P(A)$, particularly the tails of
$P(A)$,  for any physical quantity $A$ we require at finite temperatures.

\subsection{Simple-sampling algorithm}
A standard method for estimating the distribution $P(A)$ in
equation~(\ref{eqn:PofA}) is to use a simple sampling MC algorithm, where
$P(A)$ is estimated as
\begin{equation}
 P(A) = \frac{1}{M}\sum_{i=1}^{M}\delta \left(A-\langle A\rangle(\mathbf{J}^{(i)})\right)
\end{equation}
with $M$ being the number of the set of $\mathbf{J}$ generated
independently from the given distribution $P(\mathbf{J})$. 
This is only histogram accumulating the thermal average $\langle
A\rangle$ for each $\mathbf{J}$. 
It is noted that in most quenched disordered systems, the thermal
average is a difficult task at low temperatures because it takes a huge
amount of time to equilibrate. The so-called extended ensemble MC
methods \cite{IbaE} such as the multicanonical method \cite{Berg} and the
exchange MC method \cite{EMC}(parallel tempering) have partially overcome
the difficulty of slow relaxation and/or have significantly reduced
equilibration time.   
This, together with recent progress of computer power, allows us to take a
large number of average over the quenched variables. A typical example of
the total number of samples $M$ would be of order of $10^3$ in recent MC
simulations of Ising spin glasses \cite{Young,Kawashima03}. Such a number is sufficient for
evaluating the typical values, the averaged physical quantities, in most
simulations. It is, however, still difficult for the simple-sampling
algorithm to estimate rare cases, i.e., the tails of the distribution of
equation~(\ref{eqn:PofA}).   Particularly, it is impossible in principle to
obtain the distribution whose value is less than $1/M$. 
 
\subsection{Importance-sampling algorithm for quenched variables}

In this work, we discuss an efficient MC method for sampling
the quenched variables. A pioneering work was given by
Hartmann \cite{Hartmann2002}, who studied a large deviation function of a
sequence alignment.  In the literature of physics, an importance-sampling
MC algorithm for the quenched variables was proposed by K\"{o}rner,
Katzgraber and Hartmann and applied to the measurement of the
ground-state-energy distribution in the Sherrington-Kirkpatrick (SK) 
model \cite{Keorner2006}. Subsequently, the method was used for estimating
the ground-state-energy distribution for the directed polymer in a
random medium \cite{Monthus2006}. An application have been proposed to the 
problem of estimating the bit-error distribution of error-correcting
codes \cite{Iba2007}. 

An importance-sampling has been proposed for efficiently sampling the
quenched variables $\mathbf{J}$ \cite{Hartmann2002,Keorner2006}. The main
idea for the importance-sampling is to enhance a probability for finding
a set of $\mathbf{J}$ which gives a rare event for $\langle
A\rangle(\mathbf{J})$ in the distribution $P(A)$  in Markov-chain MC
simulations. In reference~\cite{Keorner2006},  a guiding function 
$\tilde{P}(A)$ is introduced as a guess for the true distribution and
the quenched variables are sampled from the inverse of the guiding
function, $1/\tilde{P}(A)$ using the importance-sampling technique.  
When the guiding function $\tilde{P}(A)$ approximates well the true distribution $P(A)$
which is unknown a priori, the probability for visiting  a set of
$\mathbf{J}$ is nearly independent of the thermal average $\langle
A\rangle(\mathbf{J})$ for a given $\mathbf{J}$. This means that a
histogram of the quantity $A$ is nearly flat. The idea behind this
algorithm is the same as the multicanonical method \cite{Berg} where a guiding
function as a function of the energy is introduced, instead of the
standard Boltzmann weight, so that a resulting energy histogram becomes flat. 

The remaining problem for performing MC simulations is how to guess the
guiding function $\tilde{P}(A)$. In \cite{Keorner2006,Monthus2006}, 
the guiding function is assumed to be a modified Gumbel distribution, in
which a set of parameters is determined by fitting the data obtained by
a preliminary simple-sampling run. The method does work quite well in
the applications to the ground-state-energy distribution of the SK spin
glass \cite{Keorner2006} and the directed polymer \cite{Monthus2006}. 
However it might rely on if the true distribution is expressed
approximately by a specific chosen analytic function such as a (modified)
Gumbel function. In this work, we use rather a learning algorithm
of the multicanonical method, which is a recursion scheme proposed by
Wang and Landau \cite{WangLandau1,WangLandau2}, to  
obtain the guiding function with no assumption of the function form. 

The importance-sampling MC algorithm used in this work is defined by the
following steps. Some initial estimate is made for the guiding function
$\tilde{P}(A)$; \textit{i.e.,} $\tilde{P}(A)$ is constant or the
resultant histogram by a simple-sampling run. An initial configuration
of $\mathbf{J}$ is randomly chosen for a set of $\mathbf{J}$ according
to the given distribution $P(\mathbf{J})$. A modification factor $f$ for
the guiding function $\tilde{P}(A)$ is initially set to a sufficient large
value, \textit{e.g.,} $f=e^1\simeq 2.718281\cdots$ \cite{WangLandau1,WangLandau2}. 

\begin{enumerate}
 \item From the current $i$-th configuration $\mathbf{J}^{(i)}$, a new
       candidate $\mathbf{J}'$ is produced by replacing a subset of
       $\mathbf{J}^{(i)}$ chosen at random with new values generated by
       the given distribution $P(\mathbf{J})$. 
 \item For the new candidate $\mathbf{J}'$, the thermal average $\langle
       A\rangle(\mathbf{J}')$ is calculated by some numerical method that
       depends on the system we consider. We call this procedure inner
       loop in the sense that one has to take an average over the
       variable $\mathbf{S}$ for a given $\mathbf{J}$, 
       while the update scheme for $\mathbf{J}$ discussed here is
       referred to as outer loop. We shall discuss implementation
       of the inner loop later. 
 \item The new candidate $\mathbf{J}'$ is accepted with probability
\begin{equation}
 P_{\rm accept} = \min\left\{\frac{\tilde{P}(\langle
		       A\rangle(\mathbf{J}^{(i)})}{\tilde{P}(\langle
		       A\rangle(\mathbf{J}')}, 1\right\}, 
\label{eqn:accept}
\end{equation}
using the current guiding function $\tilde{P}(A)$ and the $(i+1)$-th
       configuration $\mathbf{J}^{(i+1)}$ set to $\mathbf{J}'$. When
       the new candidate is rejected the current one has to be counted
       again. 
 \item The guiding function $\tilde{P}(A)$ with $A=\langle
       A\rangle(\mathbf{J}^{(i+1)})$ is updated by multiplying the current 
       value of $f$  as
       \begin{equation}
	\tilde{P}(A):=\tilde{P}(A)\times f
       \end{equation}
and a histogram of the value of $A$ is incremented as 
       \begin{equation}
	H(A):=H(A)+1,
       \end{equation}
each time a configuration $\mathbf{J}$ with the quantity $A$ is visited
       in simulations. 
\end{enumerate}

These steps continue until the accumulated histogram $H(A)$ is
approximately flat. In practice, the histogram is regarded as
{flat} when the desired range of $A$ is not less than certain
percent, say 90\%, of the average value of $H(A)$ over the range. 
Then, the value of the modification factor $f$ is reduced with
$f^{(n+1)}:=\sqrt{f^{(n)}}$ and the 
histogram is reset to $H(A)=0$ \cite{WangLandau1,WangLandau2}. Again, the
importance-sampling MC algorithm 
is performed with the new value of $f$. The guiding function
$\tilde{P}(A)$ is adjusted by gradually changing the modification factor
$f$ during the simulation so that the configurations $\mathbf{J}$ with
the value of $A$ are visited with equal probability. 
This process is repeated for many times until the value of $f$ is very
close to unity, e.g., $f=1.0001$. 

In an early stage of simulation with a large $f$, the detailed balance
does not hold precisely since the transition probability of the Markov
chain and the acceptance  probability for the update in
equation~(\ref{eqn:accept}) are modified by the factor $f$ in simulations. 
After the factor $f$ is sufficiently close to one through the iteration,
the detailed balance is recovered and the stationary distribution of the
Markov chain becomes the inverse of the guiding function,
$1/\tilde{P}(A)$.  Then, the desired distribution  $P(A)$ is estimated by 
\begin{equation}
 P(A) \propto \tilde{P}(A)H(A). 
\label{eqn:reweight}
\end{equation}  

\section{Griffiths singularity in randomly diluted Ising model}
\label{sec:Griffiths}

\begin{figure}
\includegraphics[width=18pc]{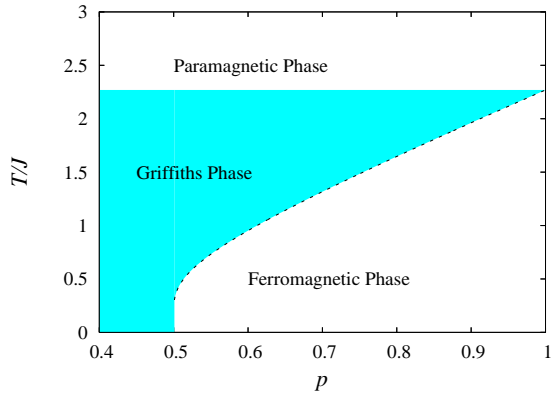}\hspace{2pc}
\begin{minipage}[b]{18pc}
\caption{
Phase diagram of the two-dimensional bond diluted Ising ferromagnet in
 the $p$-$T$ plane. 
The shaded region is called the Griffiths phase.
\label{fig:pd-2id}
}
\end{minipage}
\end{figure}

The bond-diluted Ising model is defined by the Hamiltonian,
\begin{equation}
{\cal H}(S;J) = -\sum_{\langle ij\rangle} J_{ij} S_i S_j - h\sum_iS_i
\end{equation}
where $S_i (i=1,\cdots, N)$  denote Ising spins which take $\pm 1$, 
the interactions between the spins $J_{ij}$ are independent random
variables taking $J$ and $0$ with probability $p$ and $1-p$,
respectively and $h$ represents an external field. 
The summation of the first term is over all the nearest neighbor
pairs. For $p$ close to 
one, a ferromagnetic phase transition occurs at a finite temperature
$T_c(p)$. The transition temperature decreases with $p$ decreasing and
eventually it vanishes at the percolation threshold $p_c$. The phase
diagram of the two-dimensional bond-diluted Ising model is shown in
figure~\ref{fig:pd-2id}.

According to the argument by Griffiths \cite{Griffiths}, the free energy
of the random 
bond Ising ferromagnet must be a non-analytic function of the external
field $h$ in the temperature regime above the transition temperature
$T_c(p)$ and below $T_c(p=1)$ of corresponding non-diluted system. 
The free energy has an essential singularity as a function of $h$ in the zero $h$ limit,
which is refereed to as Griffiths singularity. Physically, the Griffiths
singularity is related with the existence of arbitrarily large compact
clusters in which most of the interactions take $J$. The probability of
the existence of such large clusters is suppressed exponentially in their
size and the contribution of such clusters to the susceptibility is
proportional to the size at low temperatures. This means that the
arbitrarily large clusters have rare but significantly important effect
leading to the Griffiths singularity. However, the singularity of the
free energy is believed to be too weak to find in experiments and
numerical simulations. Meanwhile, such large clusters have also
important consequences for dynamical 
properties because they flip from one to the other stable states very
slowly at low temperatures. In fact, many
theoretical \cite{Dhar88,Bray87,Bray88} and
numerical \cite{Takano89,Colborne89}  studies have found that time
correlation functions in the temperature 
regime $T_c(p)< T< T_c(p=1)$ exhibit non-exponential slow
relaxation which is not expected in the simple paramagnetic phase. 
Thus, the temperature regime lying between $T_c(p)$ and $T_c(p=1)$ is
called the Griffiths phase. 

\begin{figure}[b]
\includegraphics[width=\textwidth]{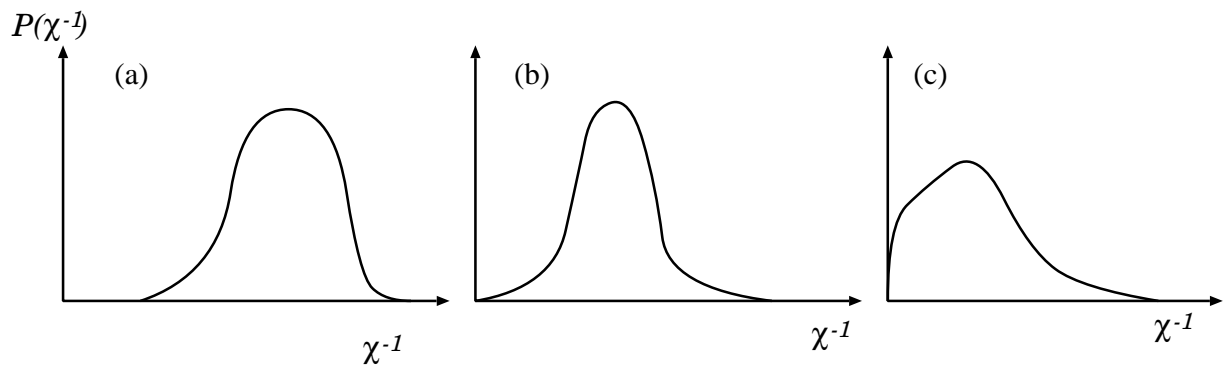}
\caption{
Schematic pictures of the distribution of the inverse susceptibility for
 a random bond Ising model. 
(a) At $T>T_c(p=1)$, the distribution is bounded. 
(b) In the Griffiths phase at $T_c(p)<T<T_c(p=1)$, the smallest edge of
 the distribution reaches the origin. 
(c) At the transition temperature $T=T_c(p)$, the statistical weight
 near the origin is significant and the average of the susceptibility
 diverges. 
\label{fig:pofis}
}
\end{figure}

In this work, we study the distribution of the (inverse) susceptibility
$\chi$ at finite temperatures. We consider that the distribution of the
susceptibility could be used as an indicator of the Griffiths
singularity on the analogy of the Bray-Moore argument \cite{BrayMoore82}
in which the density of states of the inverse susceptibility matrix is
discussed.  As is sketched in figure~\ref{fig:pofis},
the distribution of the inverse susceptibility $P(\chi^{-1})$ is bounded
at high temperatures and as the temperature decreases the edge of the
distribution touches the origin without statistical weight. We regard
this feature of the distribution as the Griffiths singularity. Using the
MC algorithm described in the previous section with setting $\langle
A\rangle=\chi^{-1}$, we evaluate the 
distribution of the inverse susceptibility numerically near the origin
in the Griffiths phase.

\section{Application of the importance-sampling algorithm}
\label{sec:Application}

We perform the importance-sampling MC simulation 
for sampling the interaction bonds $\mathbf{J}$. In
the inner loop of the step (ii) in section~\ref{sec:Algorithm} we have to
calculate the bulk susceptibility for a given $\mathbf{J}$, which is
expressed in disordered phase as 
 \begin{equation}
  \chi(\mathbf{J}) = \frac{1}{N}\left\langle
		      \left(\sum_iS_i\right)^2\right\rangle. 
 \end{equation}
For unfrustrated spin systems, cluster MC algorithms are known to be
quite efficient for reducing critical slowing down near phase
transitions \cite{SwendsenWang87,Wolff89}. Furthermore, some physical
quantities can be calculated using an improved estimator expressed in
terms of clusters, which makes
statistical error reduced substantially. The improved estimator for the
susceptibility is given by the mean squared cluster size in the
Swendsen-Wang cluster algorithm. 
Using the cluster MC algorithm with the improved
estimator \cite{SwendsenWang87}, the susceptibility of the model is
calculated with sufficient accuracy.

We firstly perform the simple-sampling algorithm for $10^3\sim 10^4$
independent interaction bonds and obtain a first guess for the guiding
function $\tilde{P}(\chi^{-1})$. Starting from an initial configuration of the interaction
bonds randomly set to $J$ with the probability $p$ and $0$ with $1-p$,
we choose a small percentage of the total bonds at random, and generate a
new candidate for the interaction bonds by refreshing the chosen bonds
with the same probability $p$. 
For the new candidate, we calculate the susceptibility by the method
mentioned above and accept it with the probability of
equation~(\ref{eqn:accept}), in which a working guiding function
$\tilde{P}(\chi^{-1})$ is used and also update by the step (iii) in
section~\ref{sec:Algorithm}. Because we are interested in the behavior of
$P(\chi^{-1})$ near th origin and not in $P(\chi^{-1})$ with a large
value of $\chi^{-1}$, we put bounds to the maximum value of $\chi^{-1}$
allowed in the simulation by $[\chi^{-1}]+3\sigma$ with $\sigma$ being
the standard deviation, which is roughly estimated in the preliminary
simple-sampling run. In actual simulation, a new candidate with the
value beyond the limit is rejected with probability one.

\begin{figure}
\begin{minipage}[t]{18pc}
\includegraphics[width=18pc]{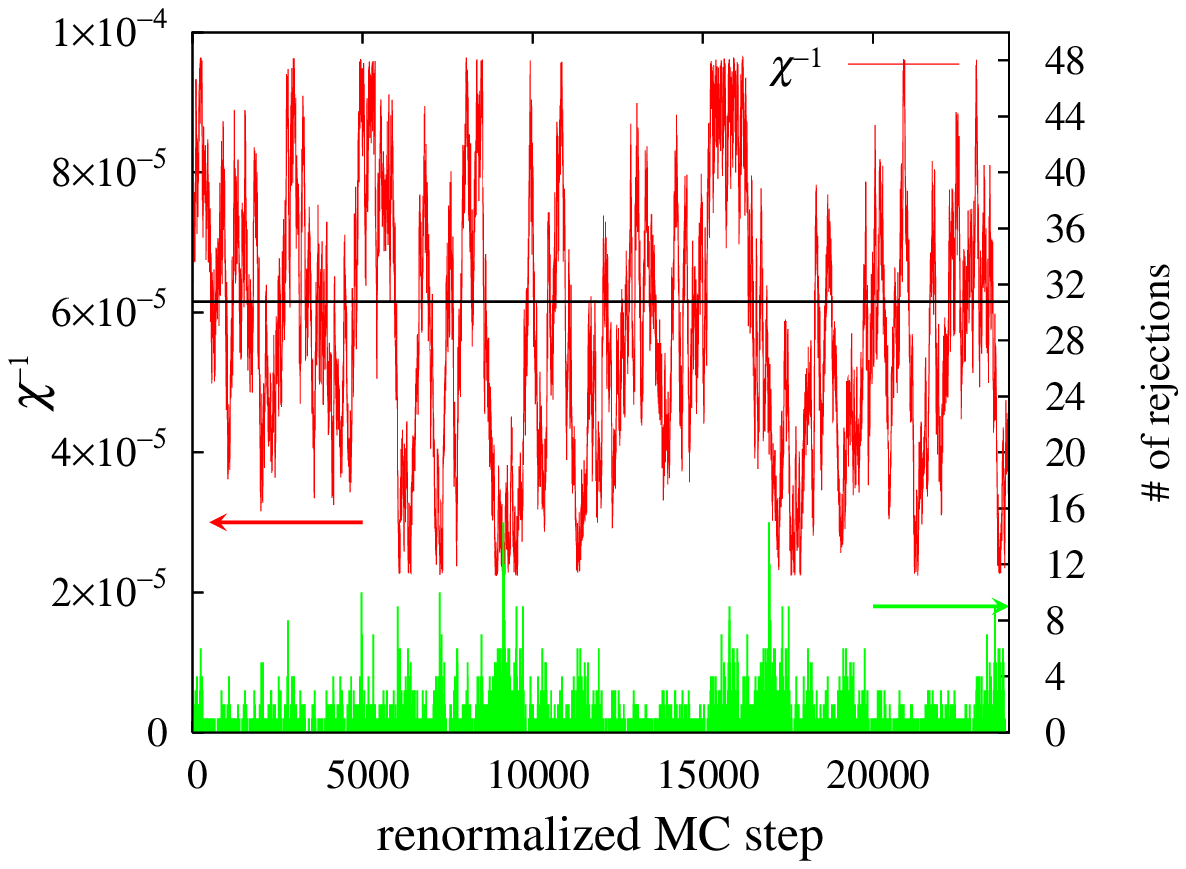}
\caption{\label{fig:history}
A Monte Carlo trajectory of the value of $\chi^{-1}$  of the
 two-dimensional bond-diluted Ising model for $L=32$, $p=0.6$ and
 $T/J=1.5$.  
The horizontal axis means renormalized MC steps which are
 incremented by one when a new value of $\chi^{-1}$ is accepted. 
The value of $\chi^{-1}$ as a function of the MC step is represented by
 the straight line and the number of rejections at the MC step is given
 by bar chart. 
}
\end{minipage}\hspace{2pc}
\begin{minipage}[t]{18pc}
\includegraphics[width=18pc]{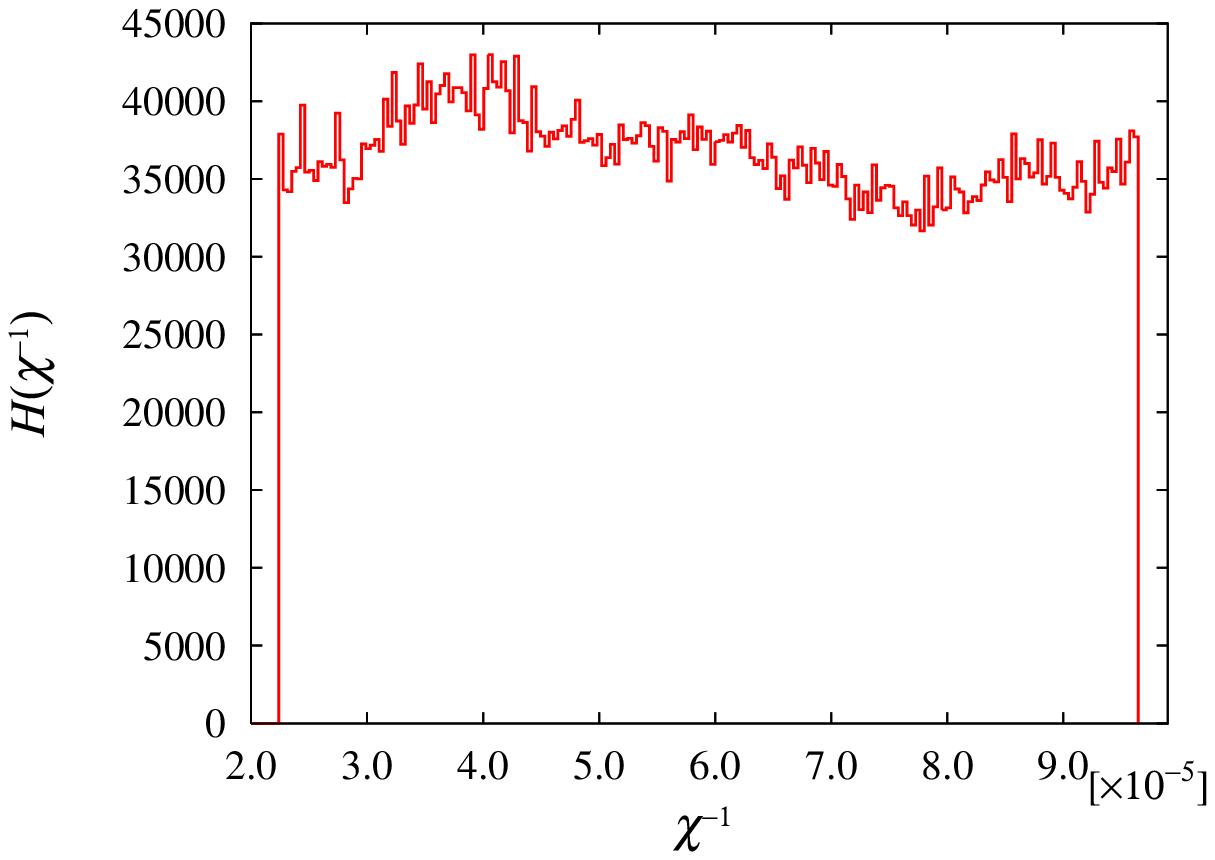}
\caption{\label{fig:histofchi}
Histogram of the inverse susceptibility obtained by an importance
 sampling algorithm of the two-dimensional bond-diluted Ising model. The
 parameters used in the simulation is the same as those in
 figure~\ref{fig:history}. 
}
\end{minipage}
\end{figure}

\begin{figure}
\begin{center}
 \includegraphics[width=24pc]{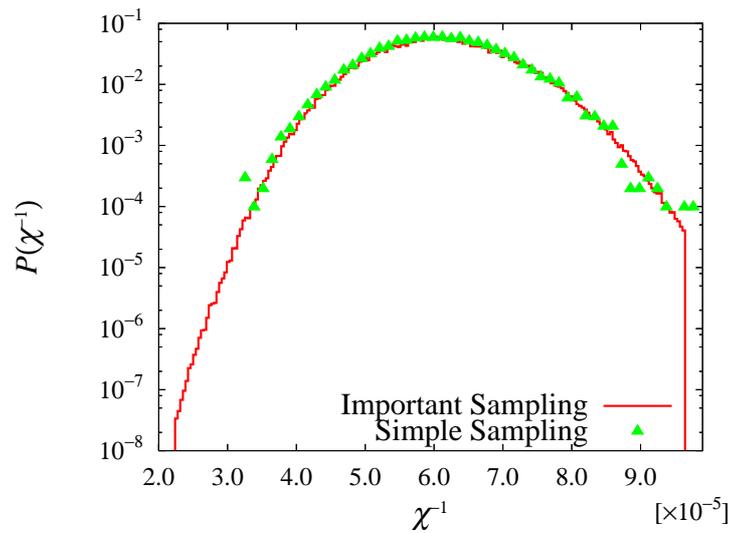}
\end{center}
\caption{\label{fig:pofchi}
Distribution of the inverse susceptibility of the two-dimensional
 bond-diluted Ising model with $p=0.6$ for $L=32$ and $T/J=1.5$. 
The solid line represents $P(\chi^{-1})$ obtained by the importance-sampling
 MC, and the triangles are that by the simple sampling. 
}
\end{figure}

Figure~\ref{fig:history} shows an example of Monte Carlo trajectory of
the value of $\chi^{-1}$ for linear size $L=32$, $p=0.6$ and $T/J=1.5$ and
figure~\ref{fig:histofchi} shows the histogram of the inverse
susceptibility in this simulation. 
It is found that the importance-sampling MC simulation can cover a wide
range of $\chi^{-1}$. 
The re-weighted distribution by the guiding function
with equation~(\ref{eqn:reweight}) and also the simple-sampling histogram for
comparison are shown in figure~\ref{fig:pofchi}.  
We see that the resolution in the simple-sampling result determined by
the number of samplings is largely improved by the use of the importance
sampling which leads to probabilities of order $10^{-8}$.


\begin{figure}
\begin{center}
\includegraphics[width=24pc]{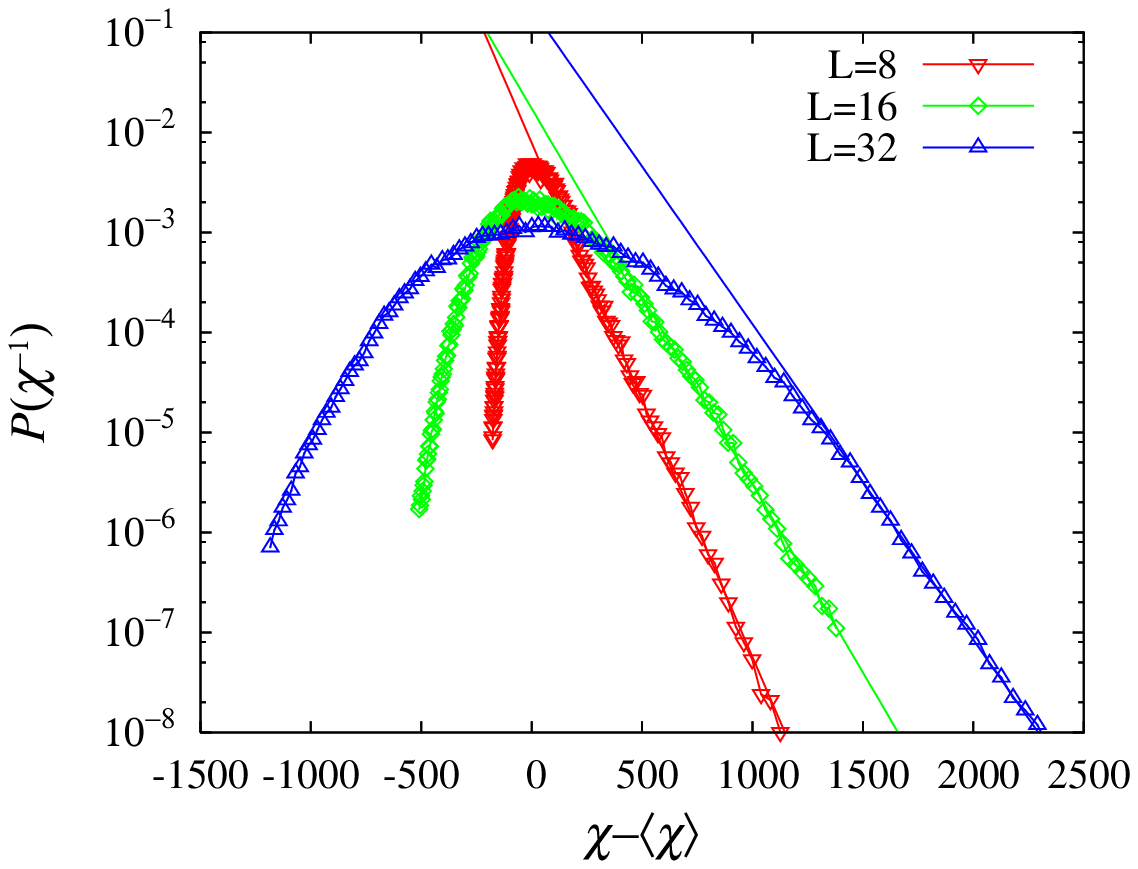}
\end{center}
\caption{\label{fig:pofchi-p060t20}
Distribution functions of the inverse susceptibility as a function of
 $1/\chi^{-1}$ of the two-dimensional bond-diluted Ising model with
 $p=0.6$ and $T/J=2.0$. The system sizes are $N=8^2(\opentriangledown)$, $16^2(\opendiamond)$ and $32^2(\opentriangle)$.
 The straight lines are the fitting result of the
 exponential function $p(\chi^{-1})=B\exp(-C/\chi^{-1})$ with $B$ and
 $C$ being fitting parameters. 
}
\end{figure}

\begin{figure}[h]
\includegraphics[width=18pc]{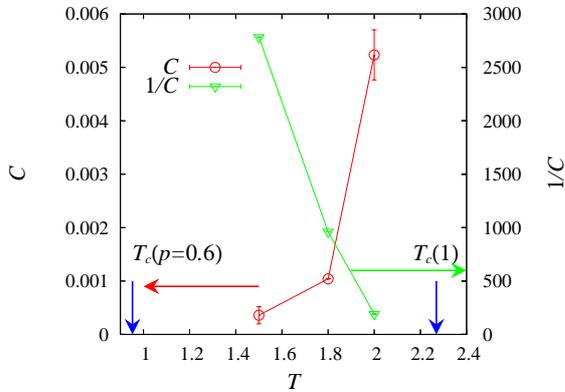}\hspace{2pc}
\begin{minipage}[b]{18pc}
\caption{\label{fig:parameterA}
Temperature dependence of the parameter $C$ for the left axis and $1/C$
 for the right.}
\end{minipage}
\end{figure}

We have calculated the inverse-susceptibility distributions  with 
$L=8$, $16$, $32$, (and $64$ for some cases) at a few
temperatures in the Griffiths and paramagnetic phases. 
In the paramagnetic phase above $T_c(p=1)$, the distribution functions
of the inverse susceptibility are scaled well as a function of
$X=(\chi-\chi_0)/\sigma$ with $\chi_0$ and $\sigma$ being the average
and the standard deviation, respectively(not shown here), and the scaling
function is well described as a Gaussian.  
Figure~\ref{fig:pofchi-p060t20} shows the distribution as a function
of $1/\chi^{-1}$ in the Griffiths phase with $p=0.6$ and $T/J=2.0$. 
In contrast to that in the paramagnetic phase, the distribution has an
exponential tail for large $\chi$  
\begin{equation}
 P(\chi^{-1})\sim \exp(-C/\chi^{-1}), 
\label{eqn:tail}
\end{equation}
where $C$  is a temperature dependent parameter.
This behavior is also described by the exponential tail of a Gumbel
distribution. The parameter $C$ is obtained by fitting the function
form of equation~(\ref{eqn:tail}) for large $\chi$ to the data for each size
and temperature. The fitting result for $C$ has still size
dependence. We thus extract the thermodynamic value by fitting a
polynomial function of $1/L$ to the value of $C$. 
The temperature dependent of the extrapolated value of $C$ for the exponential tail is shown in 
figure~\ref{fig:parameterA}. In general \cite{BrayMoore82}, we expect that
the parameter $C$ vanishes as $T$ goes to $T_c(p)$ from above since
$P(\chi^{-1})$ should be a power law of $\chi^{-1}$ for $T=T_c(p)$,
while $C$ diverges as $T$ goes to $T_c(p=1)$ since $P(\chi^{-1})$ should
have a gap from the origin $\chi^{-1}=0$ for $T>T_c(p=1)$. Our result
shown in figure~\ref{fig:parameterA} is consistent with this argument
although further studies would be necessary to confirm the
conclusion. It would be interested to see the precise functional form of
$C$ near $T_c(p)$ and $T_c(p=1)$.

\section{Conclusions and perspectives}
\label{sec:conclusion}

In summary, we have developed an importance-sampling MC algorithm for
quenched disordered systems originally proposed in
references~\cite{Hartmann2002,Keorner2006}. 
The algorithm discussed is an efficient sampling method with appropriate
weight for the quenched variables, which, for instance, correspond to
the interaction bonds in random spin systems, while most of simulations
have so far used a simple-sampling method in which the quenched
variables are generated by a given probability distribution. 
Reference~\cite{Keorner2006} suggested that a learning procedure could be
simplified by introducing  a given guiding function as the weight for
sampling the quenched variables. In this work, we explicitly
formulate the learning procedure of the weight using the multicanonical
method \cite{Berg}  combined with Wang-Landau
recursion \cite{WangLandau1,WangLandau2}. Because the method explained in
this paper does not assume a priori knowledge of the weight, it could be
complementary to the method in reference~\cite{Keorner2006} for studying a
wide class of quenched random systems. 
It should be noted that the method can be applied to finite-temperature
calculation of any distribution of observables as demonstrated in this
work, while the applications are limited to ground-state-energy
distribution previously. For example, 
the method can be in principle applicable to distribution functions of
relaxation time and/or solving time for randomly constraint satisfaction
problems. 

We have applied the algorithm to evaluate the distribution of the
susceptibility in a two-dimensional bond-diluted Ising model which
concerns the so-called Griffiths singularity. We have found that the
distribution has an exponential tail for large value of the
susceptibility and that the evaluated slope of the exponential tail
exhibits significant temperature dependence in the Griffiths phase
which is consistent with a simple argument \cite{BrayMoore82}. 
It would be interesting to confirm the Griffiths singularity in
spin-glass systems. Then, the cluster MC method used in the inner-loop
calculation for taking an average over spin variables should be replaced
by an extended-ensemble based MC method \cite{IbaE} because it does not
work efficiently in frustrated spin systems such as spin glasses. This
would require relatively extensive computational time but it would be
helpful to use massively parallel simulations for the
Wang-Landau recursion in our method . 

\begin{figure}
\begin{center}
\includegraphics[width=16pc]{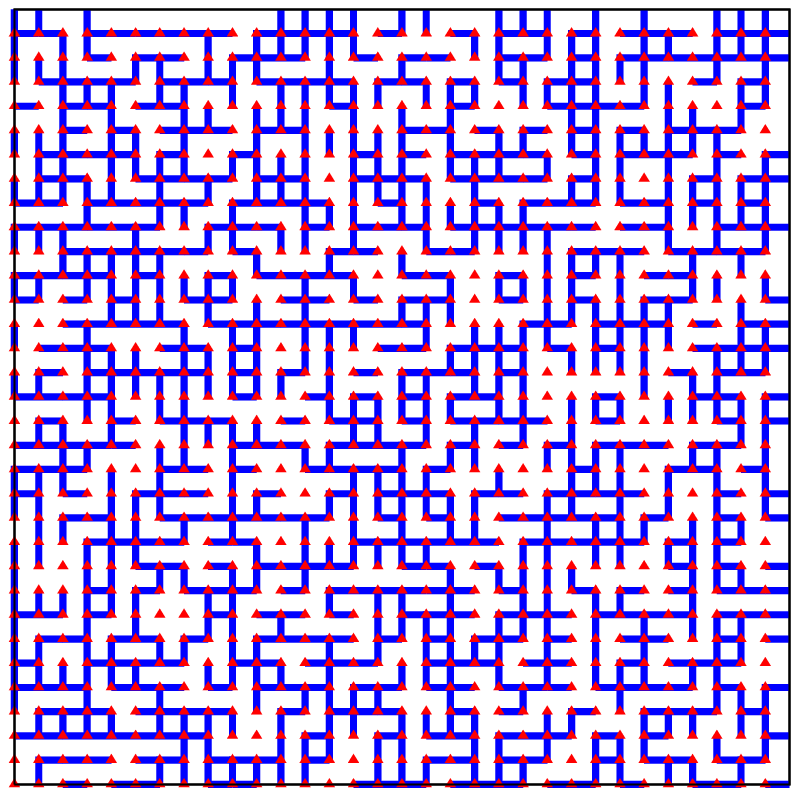}\hspace{2pc}
\includegraphics[width=16pc]{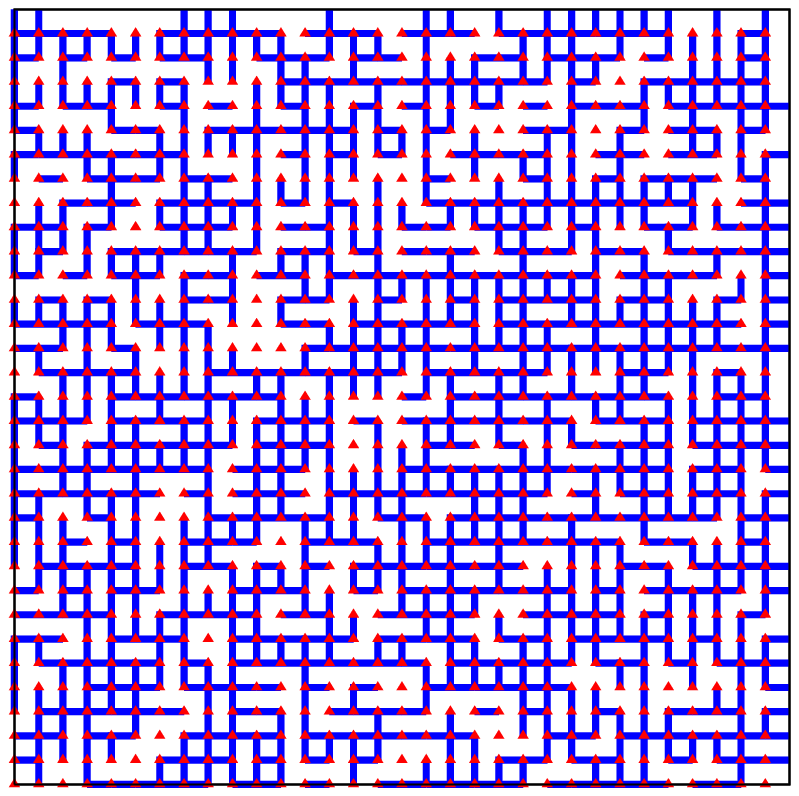}
\end{center}
\caption{\label{example1}
Examples of interaction bonds of the two-dimensional bond-diluted
 Ising model for $L=32$, $p=0.6$ and $T/J=1.5$. 
Triangles are sites of square lattice that the spins are defined on.
The blue lines represent the interaction bonds with $J$ between the
 nearest  neighbor spins.
The left viewgraph is an sample with $\chi^{-1}=9.67\times 10^{-5}$, and the right one
 shows a rare sample with the smallest value of $\chi^{-1}=2.24\times 10^{-5}$ in our
 simulations.  See also the distribution $P(\chi^{-1})$ in
figure~\ref{fig:pofchi}. 
It would be found that the right one is more dense and more compact
 than the left one.
}
\end{figure}

The importance-sampling MC method  discussed in this paper is also
regarded as a sampling method of rare events as originally pointed out
in reference~\cite{Hartmann2002}. A direct application in this respect is to
see rare samples which occur very unlikely.  Figures~\ref{example1} show
two rare examples of interaction bonds found in our simulations of the
bond-diluted Ising model. This makes possible further research on the
nature of such rare samples, for example fractal properties not discussed
here, which causes the large susceptibility and the Griffiths
singularity. This method could be generally useful as an experimental tool for
picking up the rare events and for studying their properties. 

\ack
We would like to thank A.~P.~Young and H.~G.~Katzgraber for fruitful 
discussions. 
This work was supported by the Grants-In-Aid for Scientific Research
(No.~17540348 and No.~18079004) from MEXT of Japan.
Part of the numerical simulations were partially performed on  the SGI Origin
2800/384 at the Supercomputer Center, ISSP, the University at Tokyo.

\end{document}